\documentstyle[aps, prb, epsf]{revtex}
\begin{document}

\draft

\wideabs{
\title{Mechanism of Diamond Nucleation on Titanium Substrate
under Very Low Pressure}

\author{Qijin Chen}
\address{The state Key Lab of Surface Physics, Institute of Physics, 
Chinese Academy of Sciences, Beijing 100080, China\\
Department of Physics, The University of Chicago, 5720 S. Ellis Ave., 
Chicago, 
IL $60637^\star$\\( $^\star$The corresponding address. Electronic mail:
qchen@rainbow.uchicago.edu)}

\author{Zhangda Lin}
\address{The state Key Lab of Surface Physics, Institute of Physics,
Chinese Academy of Sciences, Beijing 100080, China}

\date{\today} 
\maketitle 


\begin{abstract} 

Nucleation and its mechanism of diamond on titanium substrates under very
low pressure was studied using hot-filament chemical vapor deposition. 
Very high nucleation rates and densities ($10^8$-$10^{10}$
$\mbox{cm}^{-2}$) were obtained under 1 torr, which were 1-3 orders of
magnitude higher than the counterpart ($10^7$ cm$^{-2}$) under
conventionally low pressure (tens of torr). The effects of substrate
temperature and methane concentration under very low pressure were also
investigated, revealing that, overly high substrate temperature leads to
a relatively low nucleation density, and that higher $\mbox{CH}_4$
concentration gives rise to a higher density and a higher rate. The
nucleation mechanism is discussed in detail. While a large amount of
atomic hydrogen creates nucleating sites, sufficient supersaturation of
carbon and/or hydrocarbon species on/near the substrate surface is the key
factor for nucleation, in competition against the rapid formation of
carbide.  Very low pressure leads to long mean free path and other
benefiting effects, and hence, is critical for rapid, high-density
nucleation. Effects of substrate temperature and $\mbox{CH}_4$
concentration are also important. This further implies that
$\mathrm{C_2H_x}$ ($\mbox{x}<6$) and CH$_4$ also contribute to nucleation,
but CH$_{1-3}$ dominates under very low pressure. The very-low-pressure
method seems to be the only candidate to make diamond deposition on
titanium films applicable.  It also sheds light on how to increase the
diamond growth rate. 

\end{abstract}

}


\section{INTRODUCTION}

Great progress has been made in recent years in the field of chemical
vapor deposition (CVD) of diamond under low pressure. As one of the most
important issues, nucleation has always been one of the main focuses of
research, as it is the first and also the critical step for diamond
synthesis. Technically, to achieve rapid, high-density nucleation is very
important for achieving uniform, high-quality diamond films. It is also
critical in the effort to achieve hetero-epitaxial diamond films.
Theoretically, the mechanism has to be clear to optimize the deposition
parameters, to guide the experiments and industrial production process.
Both aspects are important for the purpose of fully utilizing diamond,
which has so many extraordinary properties.\cite{Field} For the most
commonly used silicon substrate, high-density nucleation is no longer a
problem in either microwave-plasma CVD (MPCVD) or hot filament CVD (HFCVD)
up to date. In MPCVD, it was first solved using negative bias method by
Jeng {\it et al\/}\cite{Jeng} and later developed by Yugo {\it et
al\/}.\cite{Yugo} The highest density achieved using this method is
$10^{10}$-$10^{11}$ $\mbox{cm}^{-2}$, reported by Stoner {\it et
al\/}.\cite{Stoner} Recently, Wolter {\it et al\/} applied an ac bias to
the substrate and also obtained high-density nucleation.\cite{Wolter} In
HFCVD, high-density nucleation on Si has been achieved by Zhu {\it et
al\/} using negative bias\cite{Zhu} and by Chen and Lin using electron
emission.\cite{Chen-eee,Chen-jap} An electron-emission-enhancement (EEE)
mechanism emerged.\cite{Chen-eee,Chen-jap} In addition to the EEE method,
this problem has also been solved using a very-low-pressure
technique.\cite{Chen-tsf,Chen-prb} A density comparable to the highest
value achieved in MPCVD was reported on unscratched, mirror-polished Si
substrates, with a very rapid nucleation rate and a high
uniformity.\cite{Chen-prb} Apart from these technical developments, the
mechanism of diamond nucleation is still not very clear, although models
have been postulated to accommodate those specific
techniques.\cite{Yugo,Stoner,Chen-prb,Jiang}

   Titanium, in addition to Si, is another important substrate material in
diamond synthesis due to its special industrial applications; it has been
one of the major materials for the speaker's vibrational membranes. The
super-high Young's modulus ($1.05\times 10^{12}$ N/$\mbox{m}^2$) and the
great acoustic velocity (18.5 km/s) of diamond make it alluring to further
coat thin Ti films with thin diamond films to achieve high-fidelity
acoustics. Unfortunately, Ti as a substrate has not received as much
attention as Si.  In result, diamond deposition on Ti, especially on thin
Ti films, has not been intensively studied. 

Two problems make it difficult to deposit diamond films on Ti. First, the
Ti substrate undergoes serious hydrogenation and embrittlement in the
presence of hydrogen at high temperature during the deposition process. 
At high temperature, titanium absorbs a great amount of hydrogen to form
titanium hydride.\cite{Shih,Numakura} This problem is usually overlooked
when thick Ti substrates are used (0.5 mm or thicker). However, for very
thin Ti substrates, e.g., tens of microns for a typical speaker's
membrane, this becomes a predominant problem. Hydrogenation can render the
thin Ti wafers very fragile and unusable. The other serious problem lies
in the difficulty of getting rapid, high-density nucleation because of the
easy formation of very thick intermediate TiC layers (We use TiC to denote
various forms of titanium carbide in this paper).\cite{Park} This worsens
the hydrogenation problem by prolonging the deposition duration. Park and
Lee\cite{Park} reported that nucleation began only after the TiC layers
grew to as thick as 50 $\mu\mbox{m}$, and that diamond films became
continuous only after $>240$ min deposition at the substrate temperature
of 700$^\circ\mbox{C}$. Such a long time at this temperature usually
causes very serious hydrogenation for thin Ti substrates. 

The (dc and/or ac) bias methods do not seems to be able to solve the
problems, as they usually need 30-60 min or even longer for {\it in situ}
pretreatment and nucleation under temperature
650-900$^\circ$C,\cite{Stoner,Jiang,bias} which could cause very serious
hydrogenation of the Ti substrates and formation of thick TiC layers.  The
EEE method seems to be able to get high-density nucleation within a short
time. However, using this method, as well as using the bias methods,
nucleation proceeds from the edge to the center of the
sample,\cite{Chen-jap,Zhu,Stoner2} which may take a pretty long time to get
uniform nucleation across a large substrate, and thus, result in the
serious problems. All these nucleation enhancing methods seem to fail or
need to be modified to adapt to the Ti substrates. 

Recently, Chen and Lin\cite{Chen-jmr} reported a two-step procedure to
deposit diamond films on thin Ti wafers ($\sim$$40$ $\mu$m thick) with
little hydrogenation and little TiC formation. The very-low-pressure
technique was employed to get rapid, high-density nucleation ($10^9$
$\mbox{cm}^{-2}$) within 5 min. Thus the nucleation process was
dramatically shortened. Oxygen was used to lower the subsequent growth
temperature to greatly reduce the hydrogenating speed. So far, the
very-low-pressure method seems to be the only practical one for diamond
nucleation on thin Ti wafers. In order to fully utilize this technique,
optimize the experimental parameters and make clear the mechanism, it is
necessary to make clear the detailed process of the nucleation and how it
is influenced by other experimental conditions. 

Following our previous work, we report in this paper detailed study of the
process of diamond nucleation on Ti substrates with increasing nucleation
time under very low pressure (1 torr), and the influence of substrate
temperature and $\mbox{CH}_4$ concentration under such pressure.  Rapid, 
high-density nucleation on Ti was achieved within 2 min at a density of
$10^8$-$10^{10}$ $\mbox{cm}^2$. While overly high substrate temperature
led to a lower density, higher $\mbox{CH}_4$ concentration increased the
nucleation rate and density. The mechanism of diamond nucleation was
discussed in detail, revealing the critical prerequisite of high
supersaturation of carbon/hydrocarbon on the substrate surface for diamond
nucleation. While the pressure effect, which leads to a much higher
concentration of reactive hydrocarbon radicals near the substrate surface,
is critical, the temperature and the CH$_4$ concentration are also
important factors, implying that $\mathrm{C_2H_x}$ ($\mbox{x}<6$) and CH$_4$
(through decomposition into CH$_3$)also contribute to nucleation, but
$\mathrm{CH_x}$ ($\mbox{x}<4$) is the
main hydrocarbon precursor for nucleation under very low pressure. 

\section{EXPERIMENTS}

Our experiments were conducted in a typical HFCVD device, as reported in
ref. \onlinecite{Chen-jmr}. To repeat briefly, a $\phi 140$ mm and 500 mm
long fused silica tube was used as a deposition chamber. Tungsten wires
of $\phi 0.2$ mm wound into coils of $\phi 2.5$ mm were used as
filaments, whose temperature was measure by an optical pyrometer. The
substrates were polycrystalline Ti wafers in the size of $0.5\times
8\times 10$ $\mbox{mm}^3$, whose temperature was measured with a
thermocouple. The filament-substrate distance was fixed at $\sim$8 mm.
Before loaded into the deposition chamber, all substrates were scratched
with $1.0$ $\mu$m diamond powder, and then chemically cleaned with
acetone in an ultrasonic bath for 10 min. The nucleation conditions were
listed in Table \ref{tab1}. Please note that the pressure used, 1.0 torr,
was much lower than the usual pressure, tens of torr. Subsequent
growth was also allowed when necessary. The growth condition will be
mentioned where appropriate. Four groups of experiments were done to
investigate the detailed process of nucleation with increasing nucleation
time and the effects of very low pressure, substrate temperature and
$\mbox{CH}_4$ concentration.  The as-deposited samples were characterized
mainly with scanning electron microscopy (SEM), while Raman spectroscopy
was also used with necessary.  The nucleation density was measured from
the SEM photos, and the size of the nuclei was measured from the scanned
high-resolution image files of the SEM photos with a much higher
magnification in Adobe Photoshop on computer. 

\section{RESULTS}
\label{sec:results}

Fig. \ref{fig1} shows the SEM surface morphology of the Ti substrates in
experiment Group I after (a) 2 min, (b) 3 min nucleation under very low
pressure (1 torr) and (c) 3 min nucleation under 1 torr plus 10 min growth
under normally low pressure (20 torr). The substrate temperature and
$\mbox{CH}_4$ concentration for nucleation were $\sim$810$^\circ$C and 3
vol.~\%, respectively. The growth conditions were:
$T_f=2000^\circ\mbox{C}$, $T_s=780^\circ\mbox{C}$, $[\mbox{CH}_4]= 1.5$
vol.~\%, and $F=100$ sccm. The densities in Fig.\ \ref{fig1}(a) and Fig.\
\ref{fig1}(b)  were measured to be $1.5\times 10^{10}$ $\mbox{cm}^{-2}$
and $3\times 10^9$ $\mbox{cm}^{-2}$, respectively. Evidently, very high
densities were achieved under very low pressure (1 torr) within only 2
min, which was an amazingly high nucleation rate, in contrast to the
report by Park and Lee\cite{Park} and even that on a Si substrate under
normal pressure.  Usually, the nucleation density on a scratched Ti
substrate is no larger than that on a scratched Si substrate under usual
pressure, which is $10^7$-$10^8$ $\mbox{cm}^{-2}$. Our result was 2-3
orders of magnitude higher, mainly due to the very low pressure.
Subsequent growth for 10 min gave rise to a very good crystallinity of the
diamond crystallites with an average size of 0.3 $\mu\mbox{m}$, as shown
in Fig.\ \ref{fig1}(c). Raman spectrum analysis also confirmed a good
quality of the crystallites.  The density of the crystallites was
approximately the same as that in Fig.\ \ref{fig1}(b), demonstrating that
all the nuclei in Fig.\ \ref{fig1}(b) were able to grow into a diamond
crystallite. In addition, the nuclei distributed very uniformly. An SEM
image of the same sample as in Fig.\ \ref{fig1}(a) with a lower
magnification is shown in Fig.\ \ref{fig1}(d). Scratch marks of the
diamond powder are visible. In contrast with the situation for nucleation
under normal pressure, however, no preferential nucleation along the
marks can be seen. The rough surface of the substrate was believed to to
be a result of the formation of TiC layer, which was usually very loose
and porous, and on top of which nucleation occurred. As the nucleation
durations were very short, the TiC layer could not be thick. It should be
mentioned that the substrate temperature could not be exactly the same as
the processes were so short. Moreover, the scratching for different
substrates could not be identical. These might explain the difference
between the densities in Fig.\ \ref{fig1}(a) and (b).

More detailed study of the nucleation process was performed at a higher
substrate temperature. Fig.\ \ref{fig2} shows the SEM images of the
samples in Group II after (a) 0.5 min, (b) 1 min, (c) 2.5 min, (d) 5 min
nucleation, (e) 5 min nucleation plus 10 min growth, and (f) 0.5 min
nucleation plus 14 min growth. The nucleation conditions are listed in
Table \ref{tab1}.  The difference between Group II and Group I was the
temperature. The growth conditions were:  $F=100$ sccm, $p=20$ torr,
$[\mbox{CH}_4]=1.5$ vol.~\%, $T_f=2000^\circ\mbox{C}$, and
$T_s=800^\circ\mbox{C}$. A continuous development of the nucleation with
time is presented. From Fig.\ \ref{fig2}(a), 0.5 min seemed to be too
short to make the nuclei large enough to be visible, if they formed. 
After 1 min (Fig.\ \ref{fig2}(b)), the nuclei became visible under the
specific magnification of the SEM photo, although they were still very
tiny. The average size of the visible ones was approximately 0.1
$\mu\mbox{m}$. The density was approximately $6\times 10^7$
$\mbox{cm}^{-2}$. After 2.5 min (Fig.\ \ref{fig2}(c)), the nucleation
density increased to approximately $3\times 10^8$ $\mbox{cm}^{-2}$ and the
nuclei became larger, their size being $\sim$0.15 $\mu$m. After 5 min, the
density was $\sim$$4\times 10^8$ $\mbox{cm}^{-2}$, nearly the same as in
Fig.\ \ref{fig2}(c) except that the nucleus size increased to $\sim$0.3
$\mu\mbox{m}$ and some of them merged. Obviously, the density actually
attained to its final value after 2.5 min under the specific experimental
conditions. When 10 min growth was allowed following 5 min nucleation, the
density of the diamond particles remained the same, while the size grew to
$\sim$0.4 $\mu\mbox{m}$, as shown in Fig.  \ref{fig2}(e).  From 2.5 min
(Fig. \ref{fig2}(c)) to 5 min (Fig.  \ref{fig2}(d)), the size grew
approximately linearly with the nucleation time. Fig. \ref{fig2}(e) 
indicates that the deposition rate under normal pressure (20 torr) was
much lower than that under very low pressure (1 torr). Fig. \ref{fig2}(f) 
was used to determine whether or not nuclei formed within the first 0.5
min; 14 min growth was allowed after 0.5 min nucleation. The particle
density was $\sim$$4\times 10^8$ $\mbox{cm}^{-2}$, and the particle size
averaged 0.2 $\mu\mbox{m}$, revealing that nuclei actually did form even
within the first half minute, although they were too small to be visible
in Fig. \ref{fig2}(a).  In addition, a comparison of the nucleus size and
the deposition time (14.5 min in total) with those in Fig. \ref{fig2}(d)
(0.3 $\mu\mbox{m}$, 5 min nucleation only) and Fig. \ref{fig2}(e) (0.4
$\mu\mbox{m}$, 5 min nucleation plus 10 min growth) further confirms that
the deposition rate under 1 torr was higher than that under 20 torr. 

Compare with Group I, we see that the final nucleation density at  the 
substrate temperature $850^\circ\mbox{C}$ was lower than that at 
$810^\circ$C.

The effect of CH$_4$ under very low pressure was investigated in
experiment Group III.  Fig. \ref{fig3} shows the SEM images of the samples
after 2.5 min nucleation with the CH$_4$ concentration of (a) 0.35, (b) 
0.7 and (c) 6 vol.~\% while all the other conditions were the same as in
Group II, since Fig. \ref{fig2}(c)  is also considered as one of this
group. The nuclei, if any, are invisible in Fig. \ref{fig3}(a), as the
methane concentration was too low. For 0.7 vol.~\%, several nuclei, can be
seen as tiny white spots (smaller than 0.1 $\mu$m) from Fig. 
\ref{fig3}(b) , implying a density of $\sim$$10^6$ cm$^{-2}$. For high
CH$_4$ concentration in Fig. \ref{fig3}(a), the density was approximately
$8\times 10^8$ cm$^{-2}$ with an average nucleus size of $\sim$0.22
$\mu$m. While the density in Fig. \ref{fig2}(c), corresponding to [CH$_4$]
= 3 vol.~\%, was much higher than that in Fig.  \ref{fig3}(b), the density
at 6 vol.~\% (Fig. \ref{fig3}(c)) was even higher. This demonstrates that
a higher CH$_4$ concentration led to a higher nucleation rate and a higher
density, and that a too low concentration was not suitable for nucleation.

The effect of pressure was further checked in experiment Group IV, as
shown in Fig. \ref{fig4}, the SEM image of a sample after 5 min
nucleation under pressure 10 torr. The other conditions was the same as
in Group II. The nucleation density was approximately $5\times 10^7$
cm$^{-2}$ with an average nucleus size of $\alt 0.2$ $\mu$m. This is the
typical density obtained under normal pressure (tens of torr), one order
of magnitude lower than that under 1 torr with all the other conditions
being the same (Fig. \ref{fig2}(d)), not alone that in Fig. 
\ref{fig1}(a).  This also confirms that the high nucleation density in
Fig. \ref{fig2}(f) was indeed a result of the nucleation within the first
half minute. 

\section{DISCUSSIONS}
\label{sec:discussions}

Usually, under normal pressure (tens of torr), it takes 30 min or longer
to get considerable nucleation on scratched Si or Ti substrates, and the
density is no larger than $10^7$-$10^8$ cm$^{-2}$. Compare with our
results obtained under a much lower pressure (1 torr) in Fig.~\ref{fig1}
and Fig.~\ref{fig2}, it is clear that both the nucleation rate and the
nucleation density were dramatically enhanced by the very low pressure,
for which the density was 1-3 orders of magnitude higher and the
nucleation took place at a much higher rate. While this enhancement was
mainly attributable to the much longer mean free path of the gaseous
species under the much low pressure,\cite{Chen-prb} overly higher
substrate temperature gave rise to lower nucleation density, and higher
CH$_4$ concentration led to higher rate and density, which revealed more 
information about the nucleation mechanism. 

Generally speaking, various processes take place on/near the substrate
surface during nucleation.\cite{Angus} The carbon ad-atoms may diffuse
into the substrate and form carbide; the substrate atoms may also diffuse
out of the bulk, and also form carbide. These two processes make nucleation
more difficult. It requires a sufficient high amount of carbon and/or
hydrocarbon species to achieve a high supersaturation so that nuclei can
still form in spite of these two processes. There exists a competition
between the formation of carbide and diamond nuclei. In addition, the
ad-species may move along the substrate surface, as the substrate
temperature is pretty high. The larger the mobility, the more difficult
to form stable nuclei. While various phases of carbon may form, it is
necessary to select $sp^3$ (diamond) preferentially and suppress the
formation of $sp$ and $sp^2$ (various forms of non-diamond carbon). This
will require a large amount of atomic hydrogen on/near the substrate
surface, which also plays a critical role in creating nucleating sites.
Therefore, higher concentration and supersaturation of hydrocarbon
species on the substrate surface makes it easier for them to accumulate
to form nuclei, whereas higher mobility of the ad-species and higher 
diffusivity of carbon atoms hinder the nucleation process.

For Ti substrate, the diffusion of carbon and Ti atoms is more serious 
for other substrates such as Si.
As can be seen from Fig. \ref{fig1}-\ref{fig4}, the TiC layer is very
loose and porous, which makes it very easy for the carbon atoms to
diffuse into the bulk of the substrate and for the Ti atoms to come out.
Actually, carbon atoms have the highest diffusivity in TiC among all carbides,
which makes the formation of thick TiC layers extremely easy and thus the
nucleation very difficult. Under normal pressure, a long, continuous
process of nucleation and formation of TiC is observed, usually.\cite{Park} 
This makes it extremely important to achieve very high supersaturation of
carbon on the substrate surface so that a high-density, continuous layer
of diamond nuclei forms within a short time and thus the diffusion and
the formation of carbide is highly suppressed.

As discussed in detail in ref. \onlinecite{Chen-prb}, the pressure of 1
torr is 1-2 orders of magnitude lower than tens of torr. The mean free
path of the gaseous species, $\lambda$, which is inversely proportional to
the pressure, is thus increased by a factor of several tens. Moreover, the
probability for an atom or a molecule to travel a distance $x$ without
collisions, $e^{-x/\lambda}$, depends exponentially on $\lambda$.
Therefore, an increase of $\lambda$ by one order of magnitude can increase
the probability by many orders of magnitude. The reactive hydrocarbon
radicals are mainly generated in the neighborhood of the hot filament
while the deposition takes place on the substrate at a distance ($x=8$ mm
in our experiments). The mean free path can be estimated using the formula
$\lambda = k_BT/(\sqrt{2}\pi\overline{d}^2p)$, where $\overline{d}$ is the
sum of the radii of the two colliding atoms/molecules, $k_B$ is the
Boltzman constant, $T$ is the temperature and $p$ is the pressure. We take
$T\sim 1700^\circ$C as an average temperature between the filament and the
substrate. As the gas is composed mainly of molecular hydrogen, we take the
radii of H atoms, H$_2$ molecules and $\mathrm{CH_x}$ species to be
approximately 0.75 \AA, 1.0 \AA\ and 1.6 \AA, respectively.\cite{radii}
Thus, the mean free path for hydrogen atoms is estimated to be 0.13 mm and
1.3 mm under 10 torr and 1 torr, respectively.  For hydrocarbon species,
it is approximately 0.06 mm and 0.6 mm under 10 torr and 1 torr,
respectively.  Then the probability for an H atom created by the filament
to transport to the substrate will increase from $\sim$$10^{-27}$ to
$\sim$$10^{-3}$ when the pressure changes from 10 torr to 1 torr, and that
for a hydrocarbon radical will increase from $\sim$$10^{-58}$ to
$\sim$$10^{-6}$! As the radii of the gas species can not be exact, these
exponentials may vary quit a lot depending on the values of the radii
used. Nonetheless, the big difference remains within a reasonable range of
the radii. Therefore, the concentration of both the reactive hydrocarbon
radicals and atomic hydrogen is dramatically increased on/near the
substrate surface, so is the supersaturation of carbon. The increase
amount of atomic hydrogen is critical to generate a large density of
nucleating sites and guarantee that diamond is formed, not just
non-diamond carbon. This is the main reason why the nucleation is greatly
enhanced under very low pressure. 

Apart from the long mean free path, other effects of the very low pressure
also contribute. Under very low pressure in our experiments, there existed
a strong current of electrons emitted from the hot filament, which was 0.5
A or higher.  The energy of the electrons ranged from 0 up to $\sim$80 eV,
high enough to decompose the hydrocarbon species in a collision. As we
found earlier, the energetic electrons could result in an increase of the
concentration and the kinetic energy of atomic hydrogen and reactive
hydrocarbon radicals.\cite{Chen-tsf,Chen-eee} Due to the long mean free
path, more energetic, reactive hydrocarbon radicals arrive at the
substrate since the filament temperature is much higher than that of the
environment.  In addition, under lower pressure, the filament can
decompose the gas more effectively.\cite{Setaka,Katoh} These effects also
add to the supersaturation of the reactive hydrocarbon species on the
substrate surface, presenting a nucleation enhancement effect. 

Comparison of the nucleation density between Fig.~\ref{fig1} and
Fig.~\ref{fig2} reveals that higher substrate temperature leads to a
lower nucleation density. The diffusivity of a carbon atom into the
substrate or of a Ti atom out of the substrate is proportional to
$e^{-(W_d-E)/k_BT}$, where $E$ is the energy of the atom, $W_d$ is the
corresponding energy barrier against diffusion, and $T$ is the substrate
temperature, $\sim$800$^\circ$C. Suppose $W_d$ is independent of
temperature, then at higher substrate temperature, $E$ is higher and
$k_BT$ is larger, therefore, the diffusivity is larger. This then
requires a higher supersaturation of carbon to nucleate. Also higher
temperature leads to a much higher surface mobility of the ad-species,
which is similarly proportional to $e^{(E-W_m)/k_BT}$. Now $W_m$ is the
energy barrier against the movement on the surface. Presumably, it is
very small, even smaller than the kinetic energy of the ad-species.
Therefore, at higher temperature, a larger part of the carbon is used to
form carbide, and also it is much more difficult for the
hydrocarbon ad-species to conglomerate at some nucleating sites to form
stable diamond nuclei. In result, a lower nucleation density is
observed at a lower rate, as shown in Fig.~\ref{fig2}.

The CH$_4$ concentration is naively also an important factor. Obviously,
the concentration of the hydrocarbon species near/on the sample surface
is proportional to it. Fig.~\ref{fig3}(a), (b), Fig.~\ref{fig2}(c), and
Fig.~\ref{fig3}(c) show a continuous increase of the nucleation rate and
the nucleation density. A too low concentration, e.g., 0.35 vol.~\% in
Fig.~\ref{fig3}(a), is insufficient to form nuclei as the diffusivity of
carbon into the Ti substrate is too high, and all the hydrocarbon
radicals are actually converted into TiC. Higher CH$_4$ concentration
builds up a higher supersaturation of hydrocarbon on the substrate
surface, and thus enhances nucleation.

Our estimate about the mean free path seems to indicate that the pressure
effect dominate, making all other factors unimportant. However, a
comparison of Fig.~\ref{fig3} and Fig.~\ref{fig4} tells that it is not
the case. Fig.~\ref{fig3} shows a strong influence of the CH$_4$
concentration. A too low CH$_4$ concentration could not lead to high density
nucleation, although the pressure was very low (1 torr, Fig.~\ref{fig3}(a)).
On the other hand, at normal pressure (10 torr, Fig.~\ref{fig4}), high
CH$_4$ concentration could also lead to considerable nucleation, though
the density is not high. This sheds some light on the mechanism of
nucleation.

First, not all collisions result in loss of reactive hydrocarbon radicals
and/or atomic hydrogen through recombination, as the gas is mainly
composed of molecular hydrogen. The CH$_4$ concentration is typically less
than 5 vol.~\%. In addition, only a small percentage of CH$_4$ is
decomposed to $\mathrm{CH_x}$ ($\mbox{x}<4$). The total amount of atomic
hydrogen is estimated to be of the order of 1\%. Therefore, the chance for
an H atom or a $\mathrm{CH_x}$ radical to meet an H atom or
$\mathrm{CH_x}$ is reduced by a factor of $\sim$100. In addition, even
collisions between, say, two H atoms does not necessarily lead to
recombination. Thus, the effective ``mean free path'' in terms of
recombination may be over ten times larger, if we do not consider the
energy exchange in collisions. In result, the effect of very low pressure
may not be as strong as estimated earlier. Even so, the probability for an
H atom and a $\mathrm{CH_x}$ radical to get onto the substrate without
recombination is still enhanced by a factor of $\sim$$10^2$ and
$\sim$$10^5$, respectively when the pressure is lowered from 10 torr to 1
torr.

Second, some hydrocarbon radicals, such as $\mathrm{C_2H_x}$
($\mbox{x}<6$), which also contribute to nucleation and
growth,\cite{Angus,Sun,Spear,Frenklach} are actually created through
collisions between $\mathrm{CH_x}$ ($\mbox{x}\le 4$).\cite{Frenklach}
Their concentration may decrease when pressure is lowered. In addition,
some collisions may result in exchange between an H atom and a methyl
radical, such as $\mathrm{H+ CH_4\rightleftharpoons H_2+CH_3}$, which is
not necessarily unfavorable for nucleation. 

Third, apart from $\mathrm{CH_{1-3}}$, CH$_4$ also contributes to the
nucleation and growth process as it can be converted into methyl radical
in the vicinity of the substrate.\cite{Angus,Sun} Its concentration of
CH$_4$ is basically independent of its mean free path.  Carat and Goodwin
studied the change of CH$_3$ concentration near the substrate surface with
the substrate temperature under normal pressure, and showed an
considerable increase when the substrate temperature increased from
700$^\circ$C to 1000$^\circ$C, while the concentration of
$\mathrm{C_2H_x}$ remained unchanged.\cite{Carat} Four millimeters away
from the substrate, the CH$_3$ concentration was much less dependent on
the substrate temperature. At $\sim$660$^\circ$C, the concentration of
CH$_3$ near the substrate surface was very small, only about 20\% of that
at 4 mm away. At this substrate temperature, the contribution of the
substrate is small, and the CH$_3$ was mainly from decomposition of CH$_4$
by the filament. At 800$^\circ$C, the ration increased to 40\%. This shows
that considerable part of CH$_3$ is actually generated from CH$_4$ on/near
the substrate under normal pressure. This factor makes the total CH$_3$
concentration deviate from the exponential falloff versus distance from
the filament, especially in the neighborhood of the substrate surface, so
that a considerable, though poor, nucleation density is still observed
under normal pressure. However, the substrate temperature is too low to
decompose CH$_4$ effectively. Nontheless, the contribution of
$\mathrm{C_2H_x}$ and CH$_4$ (through CH$_3$) has to be assumed to explain
the nucleation behavior under normal pressure.

Last, the hydrocarbon precursors for deposition are mainly CH$_{1-3}$ and
$\mathrm{C_2H_x}$. The former is from the decomposition of CH$_4$ partly
by filament and partly by the substrate. The latter is mainly from
chemical reaction between methyl radicals during transportation from the
filament to the substrate. Both have comparable contribution to the
nucleation under normal pressure. Using the result of Carat and
Goodwin,\cite{Carat} the CH$_3$ concentration due to the filament falls
off by a factor of at least 25 for a distance of 8 mm (the
substrate-filament distance in our experiments). Under very low pressure,
the mean free path is much longer, and thus, the part of CH$_{1-3}$ due to
the filament is dramatically increased, while the other part and the
$\mathrm{C_2H_x}$ concentration do not increase. The strong effect of the
pressure reveals that CH$_4$ and $\mathrm{C_2H_x}$ do not contribute much
under very low pressure, as compared with $\mathrm{CH_{1-3}}$.  Frenklach
suggested that the H-abstraction-$\mathrm{C_2H_2}$-addition mechanism
dominates among various deposition reactions.\cite{Frenklach} However,
this does not seem to be true in our case, as lowering pressure actually
somehow decrease the concentration of $\mathrm{C_2H_x}$. In agreement with
our argumentation, Wu and Hong's recent work also suggests that CH$_3$ is
the main hydrocarbon precursor for diamond deposition, whereas
$\mathrm{C_2H_2}$ is not efficient in diamond deposition.\cite{Wu}

On the other hand, if we assume that the effect of a pressure change from
10 torr to 1 torr was approximately countered by the change of CH$_4$
concentration from 3 vol.~\% to 0.3 vol.~\%, then the effective mean free
path of $\mathrm{CH_x}$ in terms of recombination would be about 50 times
that in terms of collision. If this was true, then the probability for,
say, a CH$_3$ radical and an H atom to recombine when colliding would be
smaller than 10\%.  However, due to the inaccuracy of the radii of various
gas species and to the contribution of CH$_4$ and $\mathrm{C_2H_x}$, the
actual situation is much more complicated. 

It is worth mentioning that the nucleation seems to take place
preferentially on protruding convex features of the surface, as can be
seen from the SEM images. As suggested by Dennig and Stevenson, this may
be attributed to a minimized interfacial energy and more dangling bonds,
etc at these sites. However, the model of enhanced nucleation at sites with
concave curvature on a 2-dimensional surface by Louchev {\it et al\/}
is yet to be verified.\cite{Louchev}

In spite of all the difficulties in getting accurate information, it is 
clear that a very low pressure increases the mean free path, greatly 
increases the concentration of reactive hydrocarbon radicals and atomic 
hydrogen, and builds up supersaturation of hydrocarbon on the substrate 
surface, which is necessary for nucleation. The obvious effect of the 
CH$_4$ concentration indicates that 
$\mathrm{C_2H_x}$ and even CH$_4$ also contribute to the deposition 
process, though far less than $\mathrm{CH_{1-3}}$ does under very
low pressure.

Since this method can achieve high density nucleation at a very rapid rate,
it can greatly reduce the duration for nucleation and thus solve to a
large extent the serious problems of 
hydrogenation and the formation of thick carbide layers, as reported in
ref.~\onlinecite{Chen-jmr}. Further work is necessary to optimize the
experimental parameter to minimize hydrogenation and the formation of
carbide.

The pressure effect can be used to promote the growth rate of diamond. As
CH$_{1-3}$ is one of the main hydrocarbon precursors, a relatively lower
pressure should be used to increase the mean free path, and the
filament-substrate distance should be as short as possible, since the
probability for a methyl radical to move without collisions from the
filament to the substrate depends exponentially on both the pressure and
the distance. In this situation, the surface morphology of the resulting
films may be different from that under normal pressure, as the ratio of
concentration between CH$_{1-3}$ and $\mathrm{C_2H_x}$ is changed for the
above reasons; this ration determines which facet, (111) or (100), of
diamond to appear finally.\cite{Spear}

\section{SUMMARY}

Very rapid nucleation was obtained on titanium substrates with a high
density ranging from $10^8$ cm$^{-2}$ to $10^{10}$ cm$^{-2}$, 1-3 orders
of magnitude higher than that obtained under normally low pressure of tens
of torr. The nucleation process was studied in detail, revealing that
rapid, high-density nucleation could occur within the first half minute. 
The effects of temperature and CH$_4$ effects indicates that, in addition
to CH$_{1-3}$, $\mathrm{C_2H_x}$ and CH$_4$ also contribute in nucleation. 
On the other hand, the strong pressure effect indicates that their role is
much less important than that of CH$_{1-3}$ under very low pressure. All
these can be explained in terms of the supersaturation of carbon and/or
hydrocarbon species on/near the substrate surface, which is a prerequisite
for nucleation. There exists a competition between the diffusion of carbon
atoms to form carbide and the nucleation process. As the formation of TiC
is so easy for titanium substrate, relatively lower substrate temperature
should be used to reduce the diffusivity of carbon atom and to decrease
the rate of the formation of TiC. Furthermore, high supersaturation of
carbon is necessary to satisfy both processes at the very beginning so
that a layer of high density nuclei is formed quickly to suppress the
formation of TiC. With a considerably high CH$_4$ concentration, very low
pressure leads to an increased mean free path, exponentially increases the
concentration of atomic hydrogen and reactive hydrocarbon radicals,
especially, CH$_{1-3}$, near/on the substrate surface. While atomic
hydrogen is critical in generating a high density of nucleating sites and
guaranteeing the formation of diamond, high hydrocarbon concentration
helps to win the competition against the quick formation of carbide. This
method has solved the problems of hydrogenation and carbide formation to a
very large extent. It also implies that lower pressure and shorter
filement-substrate distance may result in greatly enhanced growth rate.


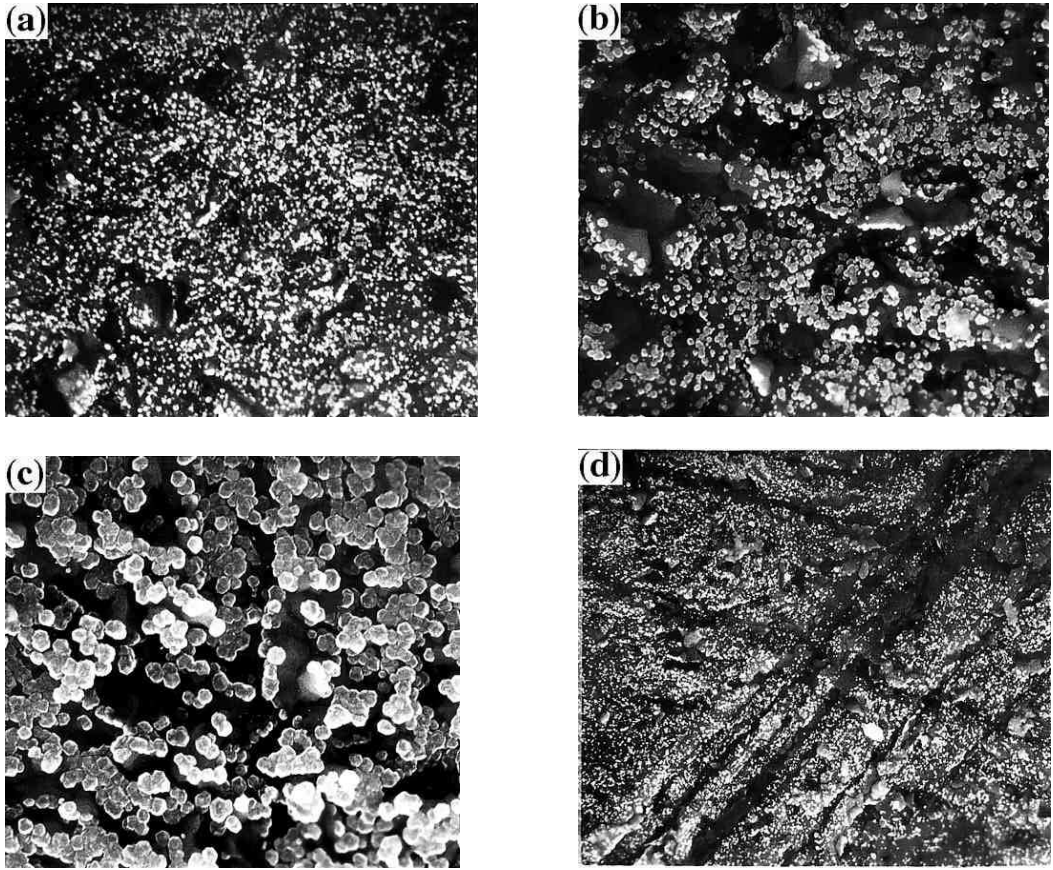
\begin{figure}
\caption{SEM images of samples in Group I after (a) 2 min, (b) 3 min nucleation
under 1 torr and (c) 3 min nucleation nucleation under $p=1$ torr plus 10
min growth under $p=20$ torr. (d) is the same as (a) but with a lower
magnification. $T_s=810^\circ$C for nucleation. Nucleation did not
occur preferentially along the scratch marks. The density  
was as high as $10^{10}$ cm$^{-2}$.}
\label{fig1}
\end{figure}
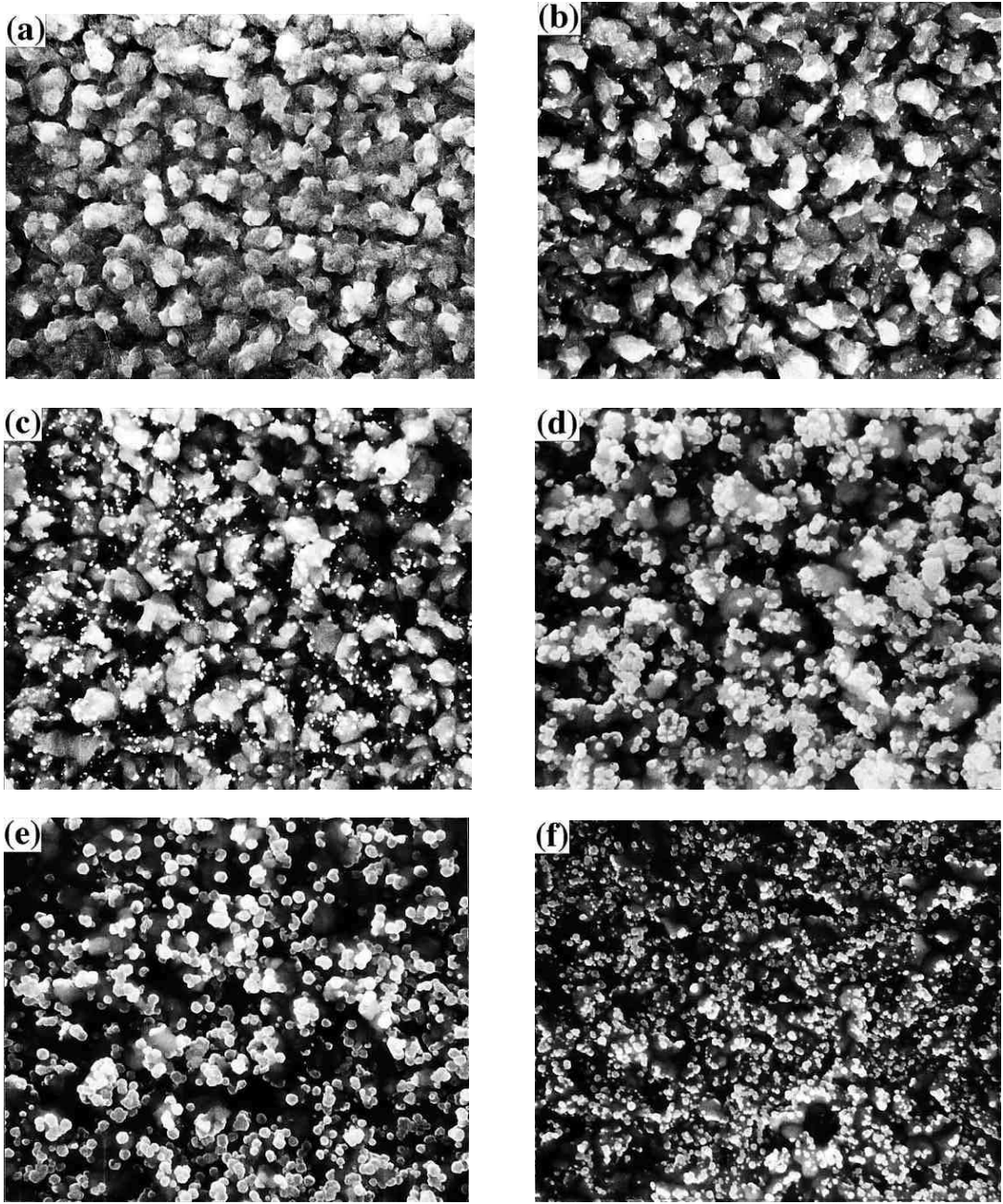
\begin{figure}
\caption{SEM images of samples in Group II after (a) 0.5 min, (b) 1 min, (c) 
2.5 min, (d) 5 min nucleation, (e) 5 min nucleation plus 10 min growth and 
(f) 0.5 min nucleation plus 14 min growth. The pressure for nucleation and 
growth was 1 torr and 20 torr, respectively. $T_s=850^\circ$C for 
nucleation. Nucleation reached its final density, $\sim$$4\times 10^8$
cm$^{-2}$, within 
2.5 min, and (f) shows that nucleation occurred during the first half minute.
The density was lower than that in Group I.}
\label{fig2}
\end{figure}
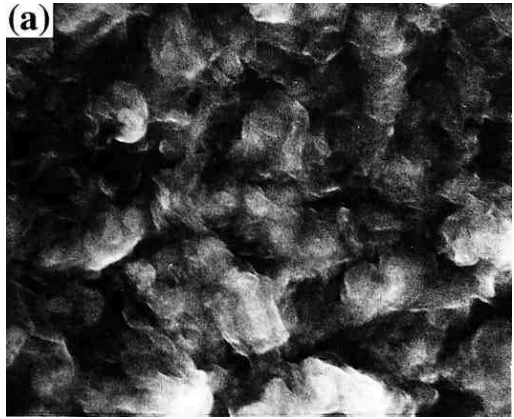
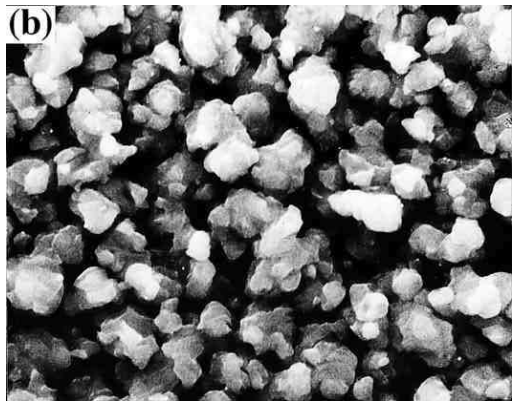
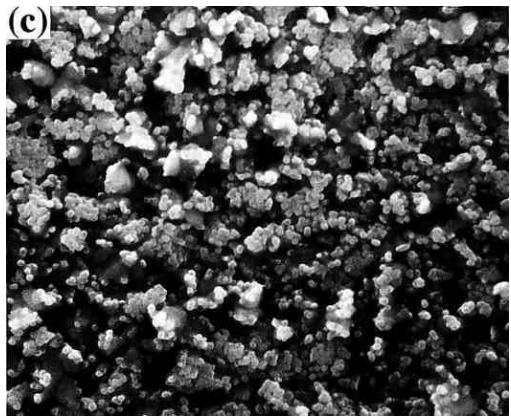
\begin{figure}
\caption{SEM photos of samples in Group III after 2.5 min nucleation under 
$p=1$ torr and different CH$_4$ concentrations: (a) 0.35,
(b) 0.7 and (c) 6 vol.~\%. Also consider Fig.~\ref{fig2}(C) as one of 
this group. The density increased with increasing CH$_4$ concentration.}
\label{fig3}
\end{figure}
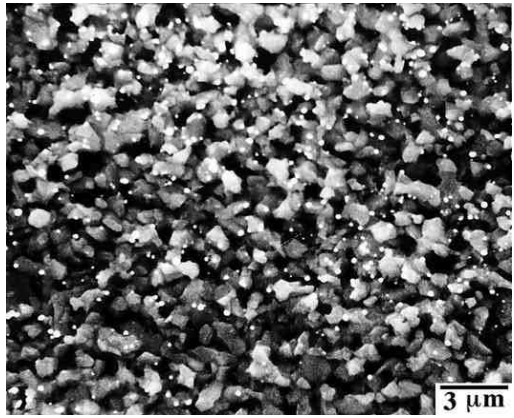
\begin{figure}
\caption{SEM image of a sample after 5 min nucleation under 10 torr. 
The density, $\sim$$5\times 10^7$ cm$^{-2}$, was one order of magnitude
lower than that obtained under 
1 torr while all other conditions were the same (Fig.~\ref{fig3}(d)).}
\label{fig4}
\end{figure}


\onecolumn
\mediumtext

\begin{table}
\caption{Experimental parameters for nucleation.}
\label{tab1}
\begin{tabular}{lcdccd}
parameters & Notations & Group I & Group II & Group III & Group IV\\
\tableline
Flow rate (sccm)\tablenotemark[1]&  $F$ & 77 & 72 & 70 &72\\
$\mbox{CH}_4$ concentr. (vol.\%) & [CH$_4$] & 3.0 & 3.0 & 0.35-6.0
&3.0\\
Filament temp. ($^\circ \mbox{C}$) & $T_f$ &2050 & 2050 & 2050
&2050\\
Substrate temp. ($^\circ \mbox{C}$) &$T_s$ &810 &850& 850 & 850\\
pressure (torr)& $p$ & 1 & 1 &1 & 10\\
duration (min) & $t$ & 2, 3 & 0.5-5.0 & 2.5 & 5
\end{tabular}
\tablenotetext[1]{sccm denote cubic centimeter per minute at STP.}
\end{table}


\begin{references}
\bibitem{radii}The radii of hydrogen and carbon atoms differ from different 
sources. Also they also depend on how they are determined. We take the 
average of their covalent radii and their Van der Waals radii for 
estimation purpose. The radii of H$_2$ molecules and $\mathrm{CH_x}$ are 
estimated based on those of H atoms and C atoms.
\end{references}
\end{document}